\documentclass[]{spie}  
\usepackage[]{graphicx}
\usepackage[]{cite}
\usepackage[]{aas_macros}
\usepackage{epsfig}
\usepackage{wrapfig}
\usepackage{bm}

\def\ba{\mbox{\boldmath$a$}}
\def\bp{\mbox{\boldmath$p$}}

\title{Simultaneous ultra-high contrast imaging and determination of time-dependent, non-common path aberrations in the presence of detector noise} 

\author{Richard A. Frazin
\skiplinehalf
{\small Dept. of Atmospheric, Oceanic and Space Sciences, University of Michigan, Ann Arbor, MI 48108} }

\authorinfo{E-mail: rfrazin {\it \_at\_} umich.edu}

\begin{document} 
\maketitle 

\begin{abstract}
Ground-based ultra-high contrast imaging, as required for direct imaging of exoplanets and other solar systems, is limited by difficulty of separating the planetary emission from the effects of optical aberrations that are not compensated by the adaptive optics (AO) system, so-called ``non-common path aberrations" (NCPAs).  Simultaneous ($\sim$ millisecond) exposures by the science camera and the AO system enable the use of ``phase diversity" to estimate both the NCPAs and the scene via a processing procedure first described by the author (R. Frazin 2013, ApJ, 767, article id. 21).   This method is fully compatible with more standard concepts used in long-exposure high-contrast imaging, such as angular differential imaging and spectral deconvolution.  Long-exposure methods find time-dependent NCPAs, such as those caused by vibrations, particularly challenging.   Here, an NCPA of the form of $\alpha \cos(k \cdot r - \omega t + \vartheta)$\ is considered.  It is shown that, when sampled at millisecond time-scales, the image plane data are sensitive to $\mbox{arg}(\alpha)$, $\vartheta$\ and $\omega$, and, therefore such NCPAs can be simultaneously estimated with the scene.  Simulations of observations with ms exposure times are reported.  These simulations include substantial detector noise and a sinusoidal NCPA that places a speckle exactly at the location of a planet.  Simulations show that the effects of detector noise can be mitigated by mixing exposures of various lengths, allowing estimation of the planet's brightness.

\end{abstract}



\section{Introduction}

The most promising method for ground-based exoplanet imaging and spectroscopy at contrast levels (planet-to-star brightness ratio) of $10^{-5}$\ or smaller combines high-order adaptive optics (AO) with a stellar coronagraph.  A stellar coronagraph is a telescopic imaging system designed to block the light from a star on the optical axis while only having minimal effect on the portion of the image surrounding the star \cite{Sivar01}.  Above the atmosphere, where the star presents the telescope with a flat wavefront, even a relatively simple coronagraph, such as the one on the Advanced Camera for Surveys on the Hubble Space Telescope, can reject the starlight with an efficiency of about $10^{-4}$.\cite{Krist05}  However, on the ground, the AO system does an imperfect job of correcting for the atmospheric distortion, leading to far lower rejection ratios,  even allowing the majority of the starlight to pass at times when the Strehl ratio is low.\footnote{The Strehl ratio is usually defined as the ratio of peak intensity intensity in the image plane divided by the peak intensity that would be seen if the incoming wavefront were perfectly flat.}  Typically, the AO system operates at visible wavelengths, while the science camera in the coronagraph captures images in the near-infrared (IR) at wavelengths of about 1.2 $\mu$m or longer.  Observing in the near-IR improves the needed contrast ratios and is required for spectroscopy of key molecules such as acetylene (C$_2$H$_2$) and methane.  The residual wavefront from the AO system, called the ``AO residual," creates speckles in the image plane that scintillate at millisecond time-scales.  If the post-AO imaging system were free of diffraction effects and optical aberrations, the amplitude of the speckle oscillations would be a smoothly decreasing function of distance from the center (until the distance corresponding to maximum spatial frequency of the AO correction is reached).  In this circumstance, a long-exposure image would have a smooth halo coming from a time-average of the stellar speckles, and subtracting it from a science image to reveal the planetary emission would be a straightforward matter.  The most common term for any post-AO aberration (i.e., an aberration that is non sensed by the AO system) is ``non-common path aberration" (NCPA).  Diffraction rings (or the equivalent, depending on the pupil geometry) and NCPAs interact with the AO residual, creating enhanced speckle amplitudes where the diffracted field is large.  This effect has been dubbed ``pinned speckle."\cite{Bloemhof04}  Thus, in a real optical system, the speckle amplitude can have a complicated structure, and to the extent that the NCPAs and pupil diffraction are unknown, the resulting long-exposure background is difficult to subtract.  Due to a variety of mechanical stresses on the telescope caused by, for example, winds and temperature gradients, the aberrations are never precisely known and vary temporally over a wide range of time-scales ranging from high-frequency vibrations to hours.  The resulting speckles have been given the name ``quasi-static speckles" due to their temporal variability, and several authors have shown that they are the major limitation in high-contrast imaging.\cite{Boccaletti04, Martinez13} Observing modes with contemporary high-contrast imaging systems, such as Project 1640, \cite{Oppenheimer12} the Gemini Planet Imager and SPHERE,\cite{McBride11,Beuzit08} involve exposure times ranging from several to tens of minutes.  In such observations, the pinned speckles appear as bright points since their time-averaged intensities are enhanced compared to the smooth background.  However, at millisecond-time scales, these speckles scintillate, providing much information about the NCPAs.

\begin{wrapfigure}{l}{0pt}
\epsfig{file=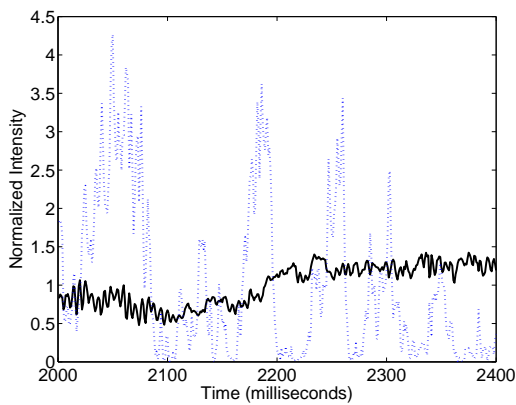,width=.4\linewidth,clip=}
\caption{\small Modulation of intensities at a single pixel of the science camera from a stellar coronagraph simulation.
The black solid line shows the time-series of the temporal variation of the planetary intensity.  The dashed blue line shows the corresponding stellar speckle intensity at the same pixel.  Both the planetary and stellar intensity are normalized to have a mean of unity in this figure.  The stellar intensity is enhanced by sinusoidal static aberration (\emph{upstream} of the coronagraph) at the spatial frequency corresponding to planet's location (from [\citenum{Frazin13}]).}
\label{fig_modulation}
\end{wrapfigure}

As the speckle overwhelms the faint planetary emission, dominating the photon noise by orders of magnitude,\cite{Racine99} post-processing of the images is necessary for ground-based high-contrast science, even with coronagraphic optics.  Since quasi-static speckles can be present over long periods of time (perhaps hours), long integration times do not significantly improve the achievable contrast.  Thus far, three types of post-processing concepts have been utilized to remove speckles from the images.  These are spectral deconvolution (SD), angular differential imaging (ADI), and locally optimized combination of images (LOCI).  SD takes advantage of the fact that the point spread function (PSF) scales with wavelength, so observations at two or more wavelengths allow for some speckle suppression.\cite{Marois00, Sparks02}.  SD is not well suited to dealing with differential aberrations that cause the PSF to loose correlation at differing wavelengths.\cite{Marois05}  ADI takes advantage of the diurnal field rotation that occurs over the course of the night when the instrument rotator is turned off (Cassegrain focus) or adjusted to maintain instrument alignment (Nasmyth focus), so that the planet appears to rotate relative to the star and the PSF.\cite{Marois06}  Thus, to the extent that speckles do not evolve as the image rotates around the pointing center, correction can be achieved.   LOCI uses the concept of a ``reference image,''  defined as any image whose subtraction from the target image would reduce the signal from the speckles while preserving the signal from the desired object.  LOCI can take advantage of reference images at different wavelengths and/or field-of-view rotations, thus incorporating the capabilities SD and ADI, as well as polarization differences.  Non-simultaneous  observations of ``naked" stars or simultaneous observations of off-axis stars may also be used to produce reference images.  (For a general discussion of LOCI see [\citenum{Lafren07}].)  Finding a good reference image to subtract is a challenge due to the temporal variability of the speckle and wavelength- (or polarization-) dependent effects, making it nearly impossible to find a single image with the same speckles as the target.  In LOCI, one starts with a set of $N$\ reference images $\{R_n\}$ and subtracts a linear combination of them, $\sum_n c_n R_n$, from the target image to get the final science image.      The ``local" nature of LOCI comes from the fact that the coefficients $\{ c_n \}$\ are chosen independently for each of a number of subregions within the target image.  In the original conception, [\citenum{Lafren07}] choose the $\{ c_n \}$\ to minimize the total squared intensity of the subregion, while the more sophisticated variant of [\citenum{Soummer12}] chooses them via Karhunen-Lo\`eve (\emph{a.k.a.}, principle components) transformation.   

The current generation of high-contrast efforts is rapidly approaching the limits of its assumptions and the level of information contained in long-exposure images.   The only major source of information that is not being utilized is the random encoding of the NCPAs by the atmosphere on millisecond time-scales.  Exploiting this source of information requires optimal processing of the data taken by the wavefront sensor (WFS) and the science camera operating in millisecond exposure mode.

\section{Information Content of Millisecond Exposures}\label{MillisecondIntro}

\begin{wrapfigure}{r}{0pt}
\begin{tabular}{rr}
\epsfig{file=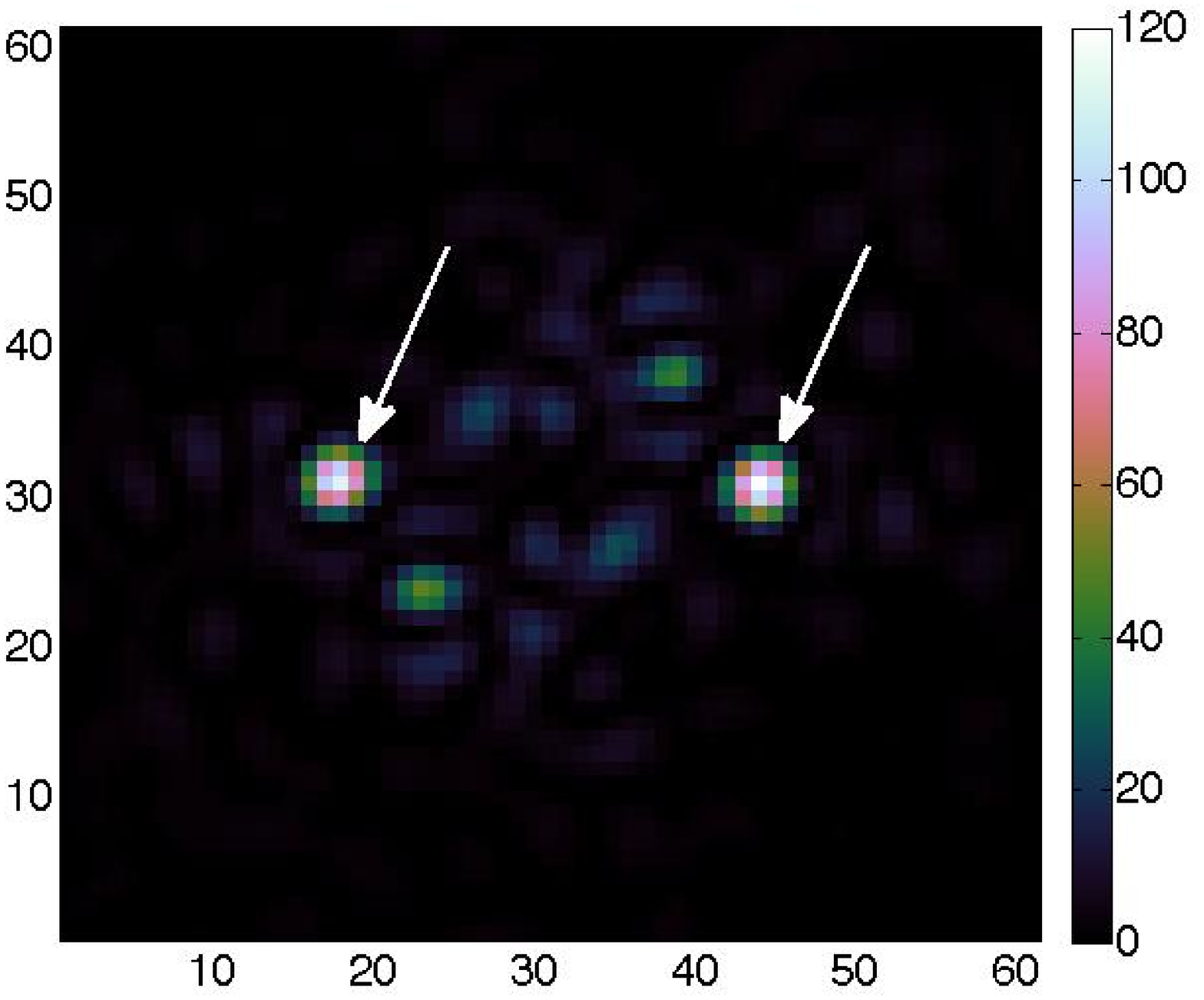,width=.3\linewidth,clip=} & 
\epsfig{file=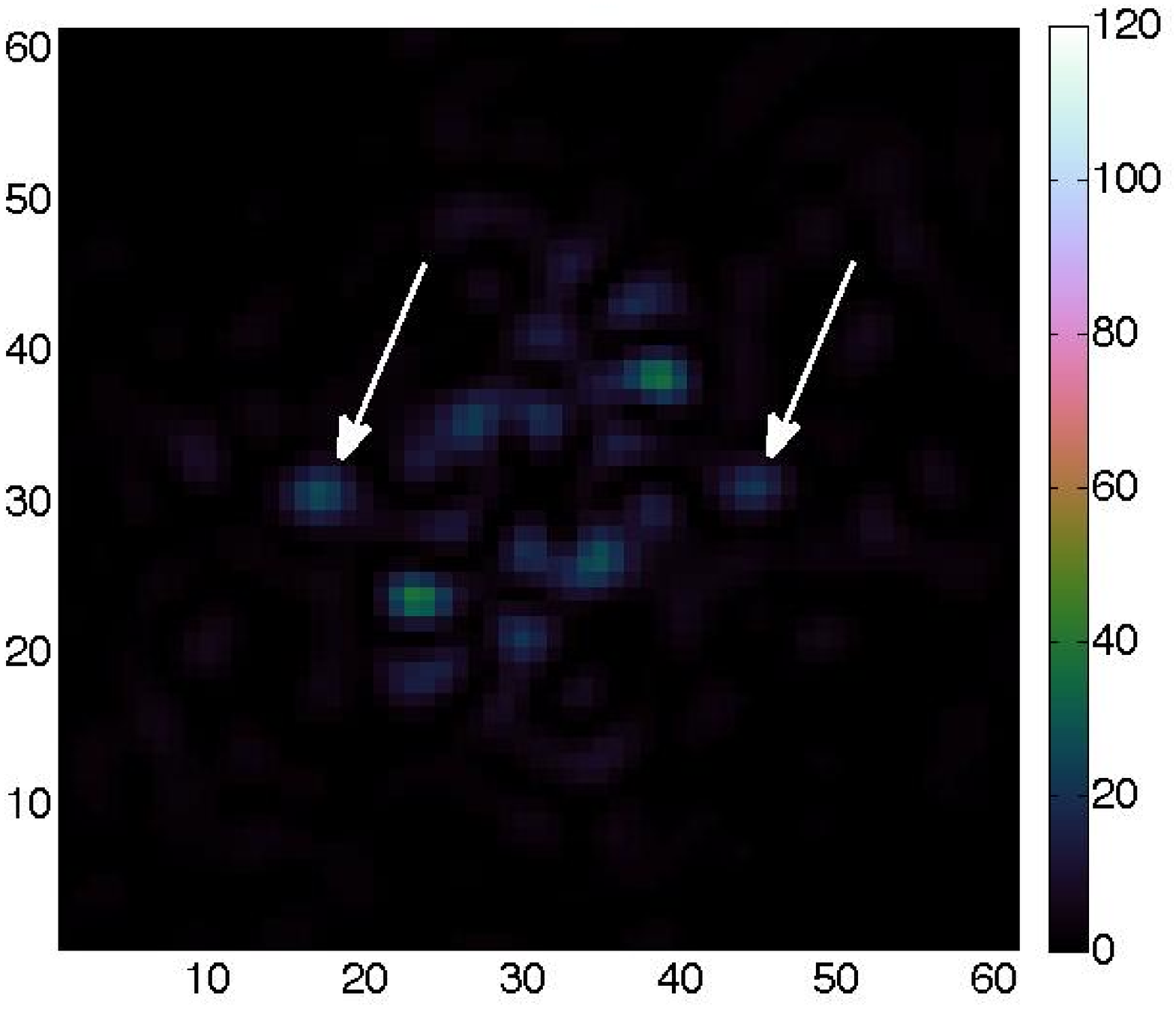,width=.3\linewidth,clip=}  \\ 
\epsfig{file=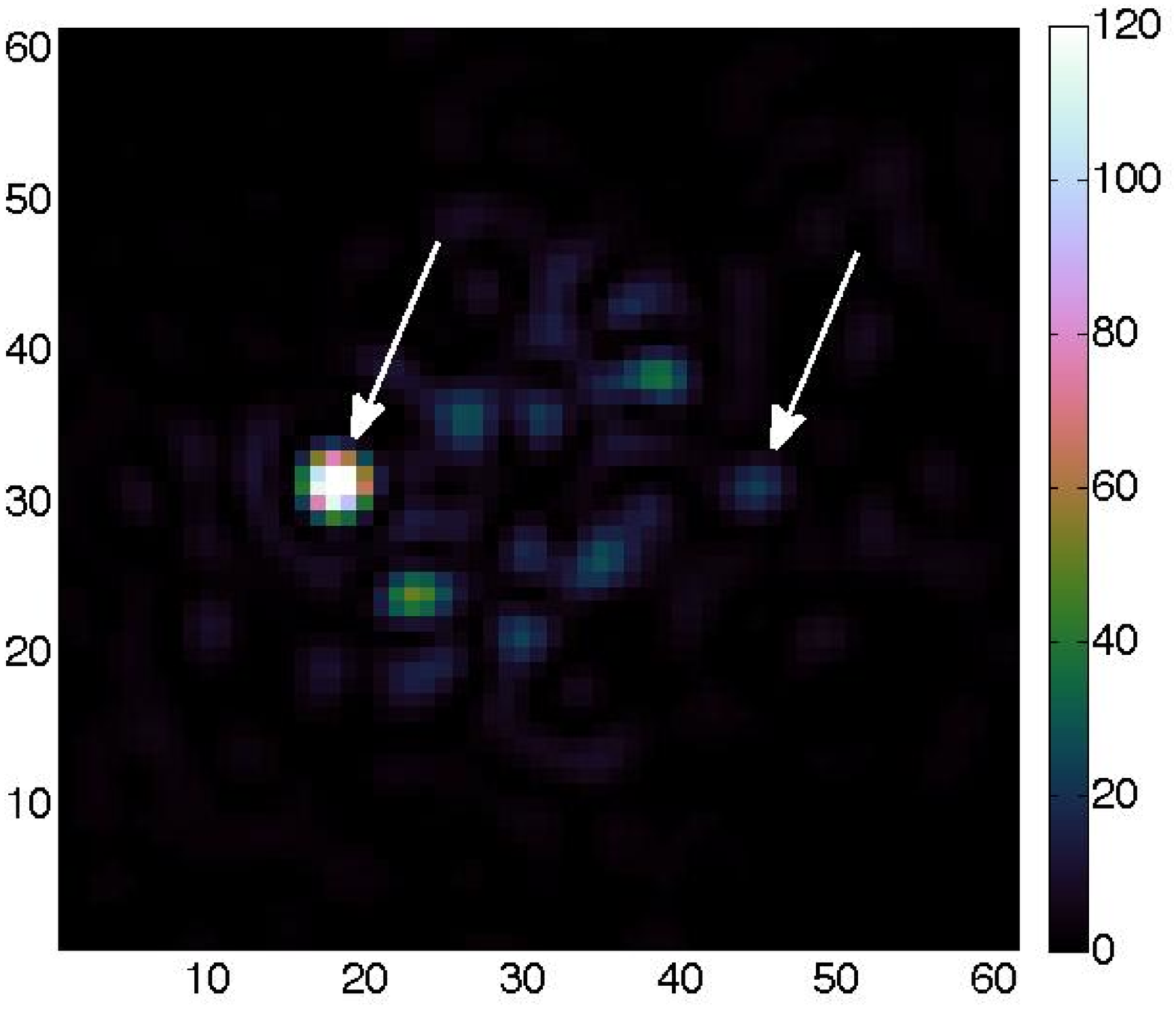,width=.3\linewidth,clip=}  & 
\epsfig{file=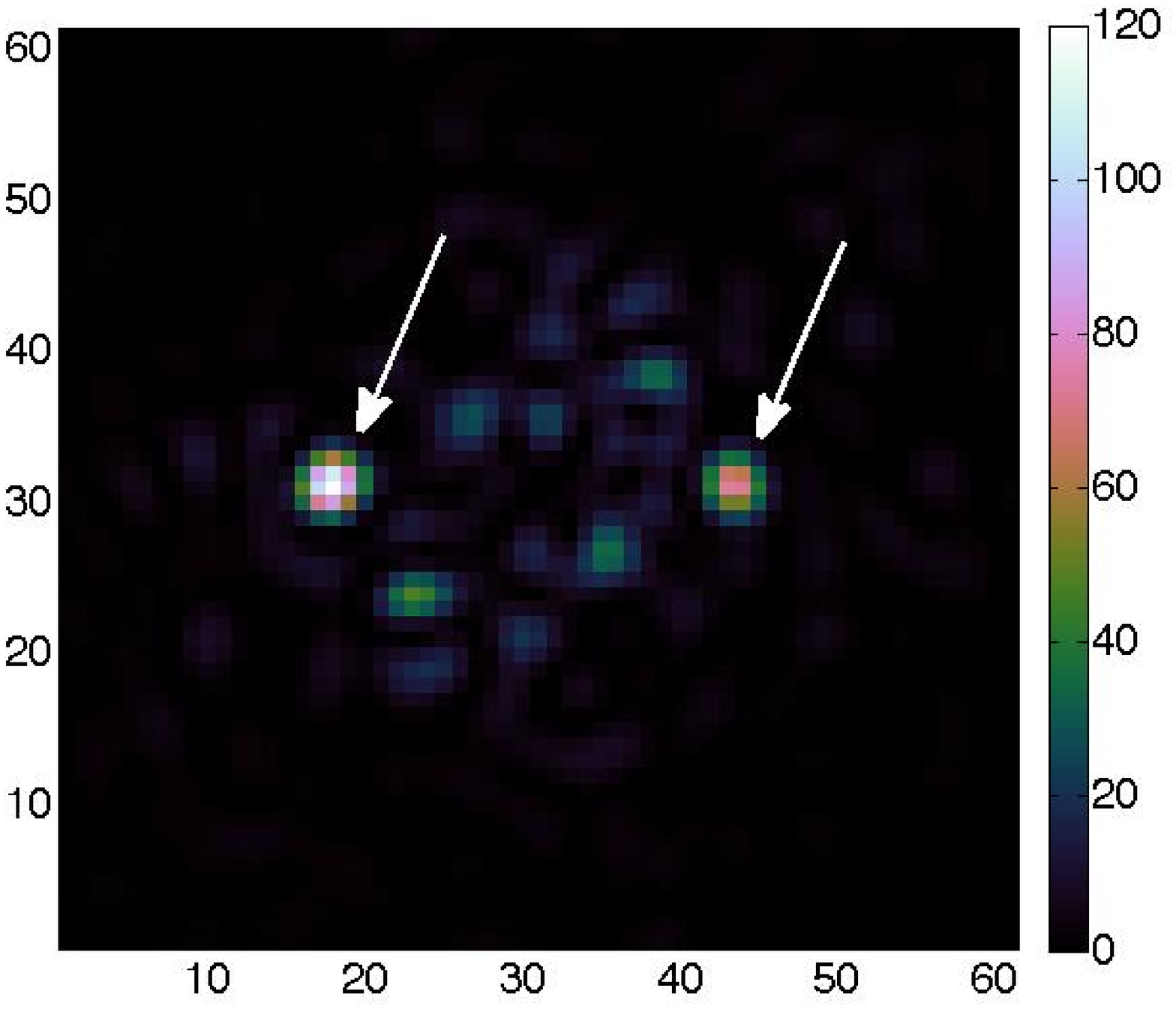,width=.3\linewidth,clip=} 
\end{tabular}
\caption{\small Stellar coronagraph simulation of a single ms science camera image showing how the instantaneous AO residual interacts with an NCPA of the form $\alpha \cos( k \cdot r + \vartheta)$.  Each panel corresponds to an NCPA with the same $k$ and $|\alpha|$, but differing in spatial phase $\vartheta$ and whether $\alpha$\ is real or imaginary.  This value of $k$\ places a speckles at the locations indicated by the arrows, at a distance of $4 \lambda/D$\ from the center.  In the \emph{top left} image, $\alpha$\ is real and $\vartheta=0$.  Similarly, the \emph{top right}, \emph{bottom left} and \emph{bottom right} correspond to $\alpha$\ values that are real, imaginary and imaginary, and $\vartheta$\ values of $\pi/2$, $0$ and $\pi/2$, respectively.  This shows that the science camera data at ms cadence are sensitive to $\mbox{arg}(\alpha)$ and $\vartheta$.  The $x-$\ and $y-$\ axes are in pixel units.  In all panels, the intensity, with units of photons/ms, is displayed.  All panels have the same color scale.}
\label{AbMod}
\vspace{-2mm}
\end{wrapfigure}
\vspace{-1.8mm}

The idea of using millisecond exposures for ultra-high contrast imaging, such as is required for exoplanet science, has been given little attention, mostly due to the fact that all of the current high-contrast efforts are based on long exposures.  The power of millisecond exposure analysis comes from two facts: \emph{1) most of the atmospheric motion is nearly frozen on that time-scale and 2) the wavefront sensor (WFS) provides a great deal of information about the AO residual}.  The key point is that at every millisecond the AO residual presents new random wavefronts.  When combined with knowledge of the wavefront,  each millisecond exposure provides more information about the NCPAs.

Using the WFS and millisecond images from the science camera has been dubbed ``random phase diversity," and it has been explored by [\citenum{Frazin13}] (henceforth, Paper I) and [\citenum{Codona13}].  In long-exposure imaging, far less information is available, as one only sees the average of all of the speckles, making the data from the WFS of little utility.   Showing a stellar coronagraph simulation result from Paper I, Fig.~\ref{fig_modulation} shows time series of the intensity seen in a single pixel of the science camera.  
The dotted curve illustrates how the AO residual modulates the stellar speckle (which, for this pixel, is enhanced by a sinusoidal NCPA) at a cadence of 1 ms.  The solid curve shows the much weaker modulation of the planetary light in the same pixel.  These two time-series, the planetary intensity and the speckle intensity, are quite different in character, with the speckle having an approximately exponential probability density function (PDF), while the PDF of the planetary intensity is somewhat localized around its non-zero mean.\cite{Gladysz10,Frazin13}  This can be understood as follows: The planetary time-series is stabilized by the AO system, as the flat part of the planet's wavefront is responsible for its intensity at this position in the image plane.  However, this star's speckle is entirely due to the random, non-flat part of the star's wavefront (the coronagraph removes the flat part) and, hence, it is much more volatile.

Long-exposure observations have difficulty distinguishing amplitude NCPAs  from phase NCPAs.  Consider pupil plane NCPA (upstream of the coronagraph) of the form $\phi(r) = \alpha \cos(k \cdot r + \vartheta)$, where $r$\ is pupil plane coordinate vector, $k$\ is the vector spatial frequency of the aberration, $\vartheta$\ is the spatial phase, and $\alpha$\ is the complex amplitude of the aberration.\footnote{Thus, the pre-coronagraph, pupil plane field is modified by this NCPA via multiplication: $E(r) \exp[ j \phi ]$, where $E(r)$\ would be the field without this NCPA. }  The imaginary part of the NCPA corresponds to an error in the wavefront amplitude and the real part corresponds to a phase error.  However, millisecond observations clearly separate these effects, as is illustrated in Fig.~\ref{AbMod}, which simulates four different sinusoidal NCPAs all using the same AO residual as input.  These four NCPAs all have same spatial frequency $k$, which places the two indicated speckles at a distance of about $4\lambda/D$\ from the center, were $\lambda$\ is the wavelength.  The top panels correspond to real NCPAs ($\alpha$\ is purely real), differing only in the phase angle $\vartheta = (0,\pi/2)$.   The bottom panels correspond to purely imaginary NCPAs ($\alpha$\ is purely imaginary), again differing only in the phase angle $\varphi = (0,\pi/2)$.  Of course, \emph{if the AO residual were precisely zero, all four of these NCPAs would lead to exactly the same two identical speckles.}  Standard long-exposure images cannot distinguish between these aberrations, although one could solve for them using well-calibrated offsets of the AO system's deformable mirror,\cite{Thomas10} in a manner similar to the procedures given in Paper I.

\begin{wrapfigure}{l}{0pt}
\epsfig{file=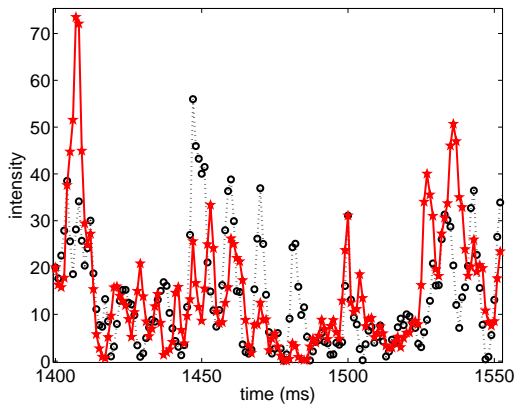,width=.4\linewidth,clip=}
\caption{\small Stellar coronagraph simulation of the effect of a vibration on a speckle in the science camera.  This speckle is caused by a vibrational NCPA of the form $\cos(k \cdot r - \omega t)$, and the modulation is caused by the interaction of the NCPA with the random AO residual. The black dotted curve corresponds to the intensity of a speckle for $\omega/2\pi =100$ Hz and the red solid curve corresponds to $\omega/2\pi = 10$Hz.  Thus, the ms data is clearly sensitive to the frequency of the vibration. \vspace{-15mm}}
\label{Vibration}
\end{wrapfigure}

Vibrations pose a particularly challenging problem to high-contrast astronomy.  Since they are sensitive to specific conditions such as wind and thermal state of the system, they are difficult to characterize with reference images.  Consider a vibration giving rise to an NCPA given by the form $\phi(r) = \alpha \cos(k \cdot r + \omega t)$, where $\omega$\ is the vibration frequency and $t$ is the time.  Clearly, at millisecond time-scales, this NCPA will manifest a modulation in the science camera that depends on $\omega$.  Fig.~\ref{Vibration} shows stellar coronagraph simulations similar to those in Paper I demonstrating that the intensity time-series of the speckle associated with this NCPA is indeed strongly dependent on $\omega$.

\section{Estimation Procedures}

Paper I used an analytical model of a stellar coronagraph and assumed knowledge of the wavefront (from the wavefront sensor) to provide a framework for simultaneously estimating the NCPAs and planetary emission.  Slightly extending this framework to include time-dependent aberrations, assume that the NCPA has a pupil plane representation of the form:
\begin{equation}
\phi(r,t) = \sum_{k=1}^{K} a_k \psi_k(r,t)  \; ,
\label{phi_expansion}
\end{equation}
where $K$\ is the number of terms required and $\{  \psi_k(r,t) \}$\ are basis functions.  Orthogonality of the  $\{  \psi_k(r,t) \}$\  is not be assumed, allowing considerable freedom in the choice of basis functions.
For convenience, the coefficients $\{ a_k \}$\ are placed into the $1 \times K$\ vector $\ba$.  Similarly, the planetary image $i_p(\rho)$, where the vector $\rho$\ is the image plane coordinate, can be expressed as a series expansion:
\begin{equation}
i_p(\rho) = \sum_{m=1}^{M} p_m \chi_m(\rho)  \; ,
\label{planet_expansion}
\end{equation}
where $M$\ is the number of terms needed, and the $\{ \chi_m(\rho)\}$\ are basis functions (they can be as simple as square pixels), and the coefficients $\{ p_m \}$\ can be placed into the $1\times M$\ vector $\bp $.  Assuming the NCPAs are small enough for the linearization
$\exp [j \phi(r)] \approx 1 + j \phi(r) $ to be valid (this should be a good approximation for aberrations that are not large enough to be immediately obvious), the measured intensity at the science camera in the image plane is given by the equation:
\begin{equation}
I(\rho,t)   =  \mathcal{A}(\rho,t) + 
\mathcal{B}(\rho,t)\cdot \ba^\star + 
\mathcal{B}^{\star \mathrm{T}} (\rho,t) \cdot \ba  +
\ba^{\star \mathrm{T}} \cdot \mathcal{C}(\rho,t)\cdot \ba +
\mathcal{F}(\rho,t)\cdot \bp + n(\rho,t) \;,  
\label{total_intensity2}
\end{equation}
where the superscripts $^\star$\ and $^\mathrm{T}$\ denote  complex conjugation and transposition respectively, $n(\rho,t)$\ represents noise in the measurment, and the $\mathcal{A}$, $\mathcal{B}$, $\mathcal{C}$\ and $\mathcal{F}$ functions depend on the AO residual at time $t$ as well as the telescope aperture function.  Paper I gives expressions for all of these functions.  $\mathcal{A}$\ is a scalar function representing speckles due to the AO residual.  $\mathcal{B}$\ is a $K \times 1$\ vector function  accounting for the interaction of the aberration $\psi_k$\ with the AO residual and represents most of the ``pinned'' speckle.  $\mathcal{C}$\  is a $K \times K$\ matrix function (which also includes for some ``pinning'') that accounts for 2\underline{nd} order interaction of the NCPAs.  Its importance depends on the amplitude of the NCPAs, but is likely to be amenable to linearization even if it cannot be taken to be negligible.  $\mathcal{F}$\ is called the ``planetary intensity kernel'' and is a $M \times 1$\ vector function describing the contribution of the planetary image to the measured intensity.   

Eq.~(\ref{total_intensity2}) forms the basis for estimation of the NCPA coefficients $\ba$\ and the planetary image coefficients $\bp$.  As $I(\rho,t) $\ is a scalar function of $\rho$\ and $t$, it must be sampled on a spatio-temporal grid, forming a three-dimensional data set, and numerical methods can be applied to estimate $\ba$\ and $\bp$, as described in Paper I.  Note that since $\mathcal{F}(\rho,t)\cdot \bp $\ represents convolution of the planetary image with the instantaneous PSF (excluding NCPAs, since they are not important for the planetary component of the image), the spatial sampling can be arranged so that this term corresponds to a Toeplitz matrix block, which can be treated with FFT-based methods.  The scale of the numerical compuation may not be small as a $64 \times 64$\ sampling of the image plane at 1000 Hz for only 1 minute yields about 246 million data values, however Kalman filters and related sequential processing algorithms should be able to reduce the computational burden. 

\section{Simulations Including Detector Noise}

One of the potential difficulties associated with millisecond exposure observations is detector readout noise.  Paper I included shot noise but did not consider detector noise, so, the first result also including both types of noise is given here.  
Since the number of detector noise counts is proportional to the number of reads, the detector noise is much less of a problem at low cadence.
Of course, long integrations greatly compromise the ability to estimate the NCPA coefficients $\ba$\ in Eq.~(\ref{total_intensity2}), so the most effective observation mode would likely combine exposures of various durations in a single estimation of the image coefficients $\bp$\ and NCPA coefficients $\ba$.   The estimation problem based on combined exposures of various durations is based on time-integration of Eq.~(\ref{total_intensity2}).  Thus, for the observations corresponding, to 1 ms, Eq.~(\ref{total_intensity2}) is integrated over 1 ms, and for a 10 s observation, the equation is integrated over 10 s.   In this way, the same formalism can be utilized for any mixture of exposure durations.  The required time-integrations require the knowledge of the AO residual from the wavefront sensor.  This avoids assuming that the time-average of the AO residual is given by a known structure function.

The simulations reported here were run using the model and procedures in Paper I with the following parameters:
The NCPAs in the simulations included 10 high-order Zernike modes and a sinusoid with a spatial frequency exactly corresponding to the location of a planet (creating a speckle there), for a total of 11 terms in Eq.~(\ref{phi_expansion}).  
\emph{Since these simulations did not include reference images, diurnal field rotation, or multiple wavelengths, estimating the correct planetary brightness with long-exposure processing would not be possible in the presence of the imposed NCPAs.}  
The separation of the planet and the star was set to $3 \lambda/D$. 
The contrast ratio of the planetary and stellar brightness was $10^{-4}$, and the stellar intensity was $10^5$ ph/ms, thus, the planetary intensity was 10 ph/ms.
The time-average of the planet's intensity at the center of its Airy disk was 0.51 ph/ms/pixel and that of the star's light at the same position was 21.7 ph/ms/pixel (note that when the NCPAs were turned off, the time average intensity of the star's speckle at that position decreased to 18.4 ph/ms/pixel, so, the NCPAs needed to be accurately estimated to have a chance to correctly estimate the planet's intensity).  
The simulations used 2000 ms exposures with phase screens (representing the AO residuals) measured by the WFS on the AEOS Adaptive Optics System.\cite{Roberts02}
 When standard deviation of the detector readout noise was set to 6.5 ph/pixel/read (corresponding to a mean of 42 photons per read),  processing all 2000 exposures gave the correct planet intensity with an error bar of 17\%.
However, integrating the first 300 ms as a single exposure (with only the noise from a single detector read in that time) and treating the remaining 1700 exposures independently (with 1700 detector reads) gave the correct planetary intensity with an error bar of 10\%.
This shows that a strategy of mixed exposure times can mitigate the effects of detector noise.  When the detector noise was turned off, the error bar decreased to 4.4\% for the mixed exposure case and 3.9\% for all-short-exposure case, as is expected, since there should be no advantage to long exposures when there is no detector noise.
In all cases the NCPA coefficients were estimated to better than 1\% accuracy.
Integrated over the image plane, the time-average stellar intensity was about $3.1 \times 10^4$\ photons/ms, meaning that, on the average, the coronagraph removed only about $69\%$\ of the starlight (the coronagraph is not effective when the wavefront is not flat).  Even though much of starlight still arrives at the science camera, the  coronagraph is still a vital piece of hardware for ground-based, high-contrast since it suppresses the zero-order diffraction and is an important source of modulation of the speckle at millisecond time-scales.

\section{Conclusion}

This paper has discussed millisecond cadence exposures in the science camera combined with simultaneous data from the wavefront sensor for the purpose of simultaneous exoplanet imaging and estimation of NCPAs.  This mode of operation has the useful potential to detect both amplitude and phase NCPAs as well as time-dependent NCPAs caused by vibrations [see Figs.~\ref{AbMod} and \ref{Vibration}].  The amplitudes of these aberrations can be estimated in the framework given in Paper I.\cite{Frazin13}  In addition, this paper gives results for simulations that include the effects of detector noise and shows that observation sequences that combine various exposure lengths can mitigate its effects.   

Millisecond imaging in the science camera may have the potential to revolutionize ground-based exoplanet imaging, but a number practical issues need to be explored first.  These include the effects of uncertainties in wavefront sensor measurements, detector noise, chromatic effects, and others, some of which are listed in Paper I.  It is important to recognize that short-exposure analysis is fully compatible with ADI and SD, as well as methods that require additional hardware to achieve phase diversity, since they can all be used in short-exposure mode.  Short exposures are also compatible with (and probably useful for) ``electric field conjugation," a technique in which the deformable mirror (DM) is adjusted in order minimize the intensity in a specific region of the image plane, thereby increasing the contrast.\cite{Thomas10}  Indeed, NCPAs determined via reference images or other types of long exposure analysis could be used a statistical priors on the estimation of the NCPAs with millisecond exposure data.  Thus, many current efforts are quite applicable to the millisecond exposure paradigm.

\acknowledgments     
 
The author would like to thank Olivier Guyon and Peter Lawson for their encouragement.


\bibliography{exop.bib}   
\bibliographystyle{spiebib}   

\end{document}